\documentstyle[preprint,aps,epsfig,axodraw]{revtex}

\def\PLB#1#2#3{{\em Phys. Lett.} {\bf B#1} (19#2) #3}

\def\PRD#1#2#3{{\em Phys. Rev.} {\bf D#1} (19#2) #3}

\def\RMP#1#2#3{{\em Rev. Mod. Phys.} {\bf#1} (19#2) #3}

\def\pb{\mbox{ pb}}

\def\gev{\mbox{ GeV}}

\begin{document}

\newcommand{\cross}{\mbox{$\rlap{\kern0.125em/}$}}

\draft

\preprint{\vbox{\hbox{USM-TH-102}}}

\title{Gauge Invariance, Color-Octet Vector Resonances and Double Technieta
Production at the Tevatron.}

\author{Alfonso R. Zerwekh$^1$ and Rogerio Rosenfeld$^2$}

\address{$^1$ {\it Departamento de F{\'{\i}}sica,
Universidad T\'ecnica Federico Santa Maria, \\
Casilla 110-V, Valpara{\'{\i}}so, Chile} \\
$^2$ {\it Instituto de F\'\i sica Te\' orica, Universidade
Estadual Paulista},\\
Rua Pamplona 145, 01405-900 - S\~ao Paulo, S.P., Brazil\\
}

\maketitle

\widetext

\begin{abstract}

We show that the usual vector meson dominance method does not
apply directly to the mixing of a color-octet vector boson
(color-octet technirho) with the gluon because of gauge
invariance. We propose a gauge invariant method where one works in
a physical basis with mass eigenstate fields. As a result, we show
that the physical technirho does not couple to two gluons,
contrary to the general belief. Consequences for the production of
a pair of color-octet, isosinglet technipions (technietas) at
Fermilab is analysed by means of a simulation of the signal and background,
including kinematical cuts. We find that the signal is too small to be
observed.

\end{abstract}

\section{Introduction}

The Standard Model has enjoyed an impressive phenomenological
success. Its agreement with the precision measurement from LEP has
increased the general confidence on the $SU(2)_L\times
U(1)_Y$ structure of the model \cite{Altarelli}. Nevertheless, the
fundamental question of how is the
electroweak symmetry broken in Nature has not been experimentally
answered yet. In fact it is generally acknowledged that the standard symmetry
breaking mechanism is  unsatisfactory from the theoretical point of view
due to the hierarchy and triviality problems. These reasons make it
necessary to investigate new signatures for alternative models of
electroweak symmetry breaking. In this paper we deal with dynamical
electroweak symmetry breaking \cite{dsb}, not committing to any
specific model but rather using some general features of realistic
models, like non-minimal technicolor.

Many non-minimal technicolor models predict the existence of
color non-singlet scalar and vector resonances. In a recent Letter
we have studied in detail the single production of a color octet
isosinglet technipion (technieta, $\eta_T$) at the Tevatron Run II \cite{us}.
In the present Letter we extend our work by considering the
production of a pair of technietas. This process was studied for the
first time by Eichten {\it et al.} \cite{EHLQ} in the context of the
one-family technicolor model and by Lane and Ramana \cite{lr} in
the context of walking technicolor. They perform calculations for the
production cross section and for the invariant mass distribution. They also
correctly point out the fact that the production of a pair of technietas is
enhanced by the mixing, in the s-channel, between a gluon and a
color octet isosinglet vector resonance, the so called ``technirho''
($\rho_{T8}$). Nevertheless, they do not take into account either
the decay of  technietas or realistic effects such as the smearing of final
momenta. We also find their description of the gluon-technirho
mixing incomplete as we shall see below.

In what follows we analyze critically the traditional description
of the gluon-technirho mixing and we emphasize its failure with a
simple example. We show how to construct a consistent gauge invariant
model using physical fields in order to describe the interactions
among technirhos, gluons, fermions and technietas. Using this
consistent approach, we perform a detailed simulation of double
technieta production at the Tevatron followed by their
decay into three jets and a photon via the process
$p\bar{p}\rightarrow \eta_{T} \eta_{T}\rightarrow \gamma g g g$.
We finally perform a simulation of the irreducible QCD background and present
our results.









\section{The Traditional Approach to gluon-technirho Mixing}

An important feature of technieta pair production is its enhancement
due to gluon-technirho mixing through vector meson dominance, as shown
in Figure 1. This approach is phenomenologically successful in describing photon-rho mixing.However, one must be careful when trying to implement the vector
dominance principle in the presence of non-abelian interactions. As an
illustrative example we consider technirho decay into two
gluons. Traditionally this process is represented only by Fig. 2a.
The technirho-gluon mixing is described by the Lagrangian

\begin{equation}
\label{eq:mix1}
  {\cal L}=-\sqrt{2}\frac{g_s}{g_{\rho}}M^2_{\rho}G^{\mu}\rho_{T\mu}
\end{equation}
\noindent
where $g_s$ is the strong coupling constant, $M_{\rho}$ is the
mass of the technirho and $g_{\rho}$ is the relevant coupling constant
of the process $\rho_T\rightarrow \pi_T \pi_T$. One estimates
$g_{\rho}$ by scaling from low energy dynamics as:

\begin{equation}
  \frac{g_{\rho}^2}{4\pi}\approx 2.97 \frac{3}{N_{TC}}.
\end{equation}
where $N_{TC}$ is the number of technicolors.
With these ingredients one obtains the usual result for the
invariant amplitude and the decay width:

\begin{equation}
-i  {\cal M}_{\mbox{a}}=-i\sqrt{2}\frac{g_s^2}{g_{\rho}}
  \frac{M^2_{\rho}}{s}f^{abc}
\left[(p_1+p_2)_{\lambda}g_{\mu \nu} + (p_3-p_2)_{\mu}g_{
  \nu \lambda} - (p_3+p_1)_{\nu}g_{\mu \lambda}
  \right]\epsilon_1^{\mu} \epsilon_2^{*\nu} \epsilon_3^{*\lambda}
\end{equation}

\begin{equation}
  \Gamma(\rho_T\rightarrow gg)= \frac{g_s^4}{g_{\rho}^2}M_{\rho}.
\end{equation}

However, the technirho is a color octet particle and a
three-point technirho vertex must exist, as shown in Fig. 2b. As
we will show in
the next section, this
self-interaction must have the same dynamical origin as the
$\rho_T\rightarrow \pi_T \pi_T$ decay and so the three-point
technirho vertex
must be proportional to $g_{\rho}$ (in fact to $g_{\rho}/\sqrt{2}$
due to normalization). As a consequence, the invariant amplitude for the
diagram shown in Fig. 2b can be written as (the extra ($-1$)
factor comes from
the addition of an extra propagator and an extra mixing term):

\begin{equation}
-i {\cal
   M}_{\mbox{b}}=i\sqrt{2}\frac{g^2}{g_{\rho}}
   f^{abc} \left[(p_1+p_2)_{\lambda}g_{\mu \nu} +
   (p_3-p_2)_{\mu}g_{\nu \lambda} - (p_3+p_1)_{\nu}g_{\mu \lambda}
   \right]\epsilon_1^{\mu} \epsilon_2^{*\nu} \epsilon_3^{*\lambda}.
\end{equation}

As we can see, when the technirho is on-shell ($s=M^2_{\rho}$) the
contribution from diagram 2b, which as far as we know has not been
taken into
account previously,  exactly cancels diagram 2a. It is an
indication that a more consistent and complete model for the
gluon-technirho interaction is needed.

\section{A consistent model for gluon-technirho interaction}

\subsection{The Gluon-Technirho Interaction}

We start by considering a general lagrangian which describes the
mixing
of two color octet vector fields

\begin{equation}
  \label{eq:L1}
  {\cal L}=-\frac{1}{4}\tilde{G}_{\mu \nu}^a \tilde{G}^{a \mu \nu}
-\frac{1}{4}\tilde{\rho}_{\mu \nu}^a \tilde{\rho}^{a \mu \nu}
+\frac{1}{2}M_G^2\tilde{G}^2
+\frac{1}{2}M_{\tilde{\rho}}^2\tilde{\rho}^2
+g_{01}\tilde{G}\tilde{\rho}
\end{equation}
\noindent
where

\begin{equation}
  \tilde{G}_{\mu
  \nu}^a=\partial_{\mu}\tilde{G}^a_{\nu}-\partial_{\nu}\tilde{G}^a_{\mu}
-\tilde{g}f^{abc}\tilde{G}^b_{\mu}\tilde{G}^c_{\mu}
\end{equation}

and

 \begin{equation}
  \tilde{\rho}_{\mu
  \nu}^a=\partial_{\mu}\tilde{\rho}^a_{\nu}-\partial_{\nu}\tilde{\rho}^a_{\mu}
-g'f^{abc}\tilde{\rho}^b_{\mu}\tilde{\rho}^c_{\mu}.
\end{equation}

We propose the following transformation laws for the fields
$\tilde{G}^a_{\mu}$ and $\tilde{\rho}^a_{\mu}$:

\begin{equation}
\label{eq:trans1}
  \delta \tilde{G}^a=-f^{abc}\tilde{G}^b\Lambda^c-\frac{1}{\tilde{g}}\partial\Lambda^a
\end{equation}
\noindent
and

\begin{equation}
\label{eq:trans2}
  \delta \tilde{\rho}^a=-f^{abc}\tilde{\rho}^b\Lambda^c-\frac{1}{g'}\partial\Lambda^a.
\end{equation}

Imposing gauge invariance we find the relations:

\begin{equation}
\label{g01mr}
  g_{01}=-\frac{\tilde{g}}{g'}M_{\tilde{\rho}}^2=-\frac{g'}{\tilde{g}}M_{G}^2
\end{equation}

Notice that from equations (\ref{eq:mix1}),(\ref{eq:L1}) and
(\ref{g01mr}) we find that $g'=g_{\rho}/\sqrt{2}$. On the other
hand, the lagrangian (\ref{eq:L1}) and the relations (\ref{g01mr})
define the following mass matrix:

\begin{equation}
  \label{eq:matrix}
  {\cal M}=\frac{M_{\tilde{\rho}}^2}{2}\left(
\begin{array}{cc}
\frac{\tilde{g^2}}{{g'}^2} & -\frac{\tilde{g}}{g'}\\
 -\frac{\tilde{g}}{g'} & 1
\end{array}
\right)
\end{equation}

It is easy to see that the matrix ${\cal M}$ is singular and so it
has a
null eigenvalue. The eigenstates can be written as:

\begin{eqnarray}
\label{eq:campfisicos1}
  G_{\mu}^a&=&\tilde{G}_{\mu}^a\cos \alpha + \tilde{\rho}^a \sin
  \alpha \\
\label{eq:campfisicos2}
\rho_{\mu}^a&=&-\tilde{G}_{\mu}^a\sin \alpha +
\tilde{\rho}_{\mu}^a \cos \alpha
\end{eqnarray}
\noindent
where the mixing angle $\alpha$ is defined by $\tan
\alpha=\tilde{g}/g'$. Remark that field $G_{\mu}^a$ represents a
physical (massless) gluon and $\rho_{\mu}^a$ is a physical (and
massive) technirho. It is instructive to note from equations
(\ref{eq:trans1},\ref{eq:trans2}) and equations
(\ref{eq:campfisicos1},\ref{eq:campfisicos2}) that the new
(physical)
fields transform according to
 \begin{eqnarray}
   \delta G^a&=&-f^{abc}G^b\Lambda^c-\frac{1}{g}\partial\Lambda^a\\
\delta \rho^a&=&-f^{abc}\rho^b\Lambda^c,
 \end{eqnarray}
\noindent
(where we have used the useful definition
$g=\tilde{g}\cos\alpha=g'\sin\alpha$ and we have
omitted Lorentz indexes). That means
that the $G_{\mu}$ field transforms as a gauge field and
$\rho_{\mu}$ as a matter field in the adjoint representation of SU(3).

Now we can analyze the $\rho_{T8}-G-G$ interaction. It can be
obtained from lagrangian (\ref{eq:L1}) and the definition of the physical
fields. The result is

\begin{equation}
  \label{eq:Lrgg}
  {\cal L}_{\rho GG} = (-\tilde{g}\cos^2 \alpha \sin \alpha +g' \sin^2
  \alpha \cos \alpha)f^{abc}
(\partial_{\mu}G^a_{\nu}G^{b\mu}\rho^{c\nu}+
  \partial_{\mu}G^a_{\nu}\rho^{b\mu}G^{c\nu}+
 \partial_{\mu}\rho^a_{\nu}G^{b\mu}G^{c\nu}).
\end{equation}
\noindent
Using the fact that $\tilde{g}\cos\alpha=g'\sin\alpha$ we can see
that
the coupling constant of the $\rho_{T8}-G-G$ interaction vanishes
exactly.

\subsection{Introducing the Technieta}

Because our main goal is to study the production of a pair of
technietas, it is necessary to include in our model the gluon and
technirho interactions with technietas. At this point we implement
the vector dominance principle by assuming that the technieta only
interacts with the field $\tilde{\rho}_{\mu}^a$. Using this
hypothesis we can write the following lagrangian:

\begin{equation}
  {\cal L}=\frac{1}{2}D_{\mu}\eta^a D^{\mu}\eta^a -\frac{m^2}{2}\eta^2
\end{equation}
\noindent
with

\begin{equation}
  D_{\mu}\eta^a=(\delta^{ab}\partial-g'f^{abc}\tilde{\rho}^c)\eta^b.
\end{equation}

After the diagonalization of the gluon-technirho mass matrix, we
arrive at

\begin{eqnarray}
  {\cal L}&=&\frac{1}{2}\partial\eta^a \partial\eta^a
  -\frac{m^2}{2}\eta^2-g'\sin\alpha f^{abc}G^c\eta^b\partial\eta^a
  -g'\cos\alpha f^{abc}\rho^c\eta^b\partial\eta^a \nonumber\\
& &+\frac{{g'}^2}{2}\sin^2\alpha f^{abe}f^{aci}G^eG^i\eta^b\eta^c
   +\frac{{g'}^2}{2}\cos^2\alpha f^{abe}f^{aci}\rho^e\rho^i\eta^b\eta^c
   \nonumber\\
& &+{g'}^2\sin\alpha\cos\alpha f^{abe}f^{aci}G^e\rho^i\eta^b\eta^c .
\end{eqnarray}

\subsection{Introducing Quarks}

In a similar way, we assume that quarks only interact with the
field $\tilde{G}_{\mu}^a$ so we write the lagrangian

\begin{equation}
  {\cal L}=\bar{\psi}i\gamma^{\mu}(\partial_{\mu}
  -i\tilde{g}\tilde{G}^a_{\mu}\frac{\lambda^a}{2})\psi-m_q\bar{\psi}\psi.
\end{equation}
\noindent

After the diagonalization of the gluon-technirho mass matrix, we
obtain

\begin{equation}
  {\cal L}=i\bar{\psi}\gamma^{\mu}\partial_{\mu}\psi -m_q\bar{\psi}\psi +
  g\bar{\psi}\gamma^{\mu}G^a_{\mu}\frac{\lambda^a}{2}\psi -
  g\tan\alpha \bar{\psi}\gamma^{\mu}\rho^a_{\mu}\frac{\lambda^a}{2}\psi .
\end{equation}

\section{Double Technieta Production}

Now we can study the production of a pair of technietas in a
consistent manner. The relevant Feynman diagrams for this process are shown
in figure 3. Note that, since the physical technirho does not couple
to two gluons, the sub-process $gg \rightarrow \eta_{T8}\eta_{T8}$ is
not resonant.

In order to perform a realistic simulation, we take into account
the technieta decay and the smearing of final momenta. We focus our
attention on the complete process
$p\bar{p}\rightarrow\eta_{T8}\eta_{T8}\rightarrow \gamma g g g$. We
chose this specific process in order to maximize the signal
over background ratio.

We use the MADGRAPH/DHELAS \cite{MAD,HELAS} package in order to
estimate the background. We include all the leading order QCD processes
that contribute to the $\gamma + 3\mbox{ jets}$ final state, where
the jets are originated by gluons and (anti-)quarks of the first two
generations only. The dominant process is $u \bar{u} \rightarrow
\gamma g g g$. We make the convolution of the parton level processes with CTEQ
structure functions \cite{CTEQ}.

During the event generation we use the following kinematical cuts:

\begin{equation}
  p_T > 15.0 \gev \;\;\; \mbox{ for photons and jets }
\end{equation}

\begin{equation}
  \left|\eta\right| < 1.5 \;\;\; \mbox{ for photons}
\end{equation}

\begin{equation}
  \left|\eta\right| < 3.0 \;\;\; \mbox{ for jets}
\end{equation}

\begin{equation}
  \left|\cos \theta_{ij}\right| < 0.95
\end{equation}
\noindent
where $\eta$ represents the pseudo-rapidity and $\theta_{ij}$ is
the angle between two jets.

In order to include the final momenta smearing effect, we use the
following detector resolution:

\begin{equation}
  \frac{\sigma(E)}{E}=\frac{0.20}{\sqrt{E(\gev)}}\;\;\; \mbox{ for
  photons}
\end{equation}

\begin{equation}
  \frac{\sigma(E)}{E}=\frac{0.80}{\sqrt{E(\gev)}}\;\;\; \mbox{ for jets}
\end{equation}

The background cross section obtained with
the cuts presented above is $\sigma_{\mbox{\tiny{back}}}=113 \pb$.
The cross section obtained for the signal is shown in Table 1, which is many
orders of magnitude smaller (we assume that all colored technipions
have the same mass).

In order to improve our analysis we implement the following cuts:

\begin{equation}
  p_T > \frac{M_{\eta}}{2}-15.0 \gev \;\;\; \mbox{ for photons and jets }
\end{equation}

\begin{equation}
M_{\eta}- \frac{M_{\eta}}{10}<M_{\gamma\mbox{\tiny{ jet}}
}<M_{\eta}+20 \gev .
\end{equation}
\noindent
and

\begin{equation}
  \left|\eta\right| < 0.7 \;\;\; \mbox{ for photons and jets}
\end{equation}
\noindent
where $M_{\gamma\mbox{\tiny{ jet}}}$ is the invariant mass of the
photon and a jet system.

Figure 4 shows, as an example, the expected event distribution for
signal and background, with $M_{\eta}=150$ GeV and $M_{\rho}=410$
GeV for the Tevatron Run II, assuming an integrated luminosity of
2 fb$^{-1}$. It also shows the improvement due to a better energy
resolution of the detectors. In all the cases we analyzed, the
resulting significance was lower than 2 and we can not expect to
observe the production of a pair of technietas at the Tevatron.

\section{conclusions}

In this work we have performed a detailed analysis of the
production
of a pair of technietas. During our study we have shown that the
traditional way to deal with the technirho-gluon mixing is
inadequate due to the non-abelian interactions felt by the technirho. This
fact led us to build a consistent model for the interactions among
technirhos, gluons, technietas and quarks. Unfortunately, when we
applied these results to the  production
of a pair of technietas we conclude that we have no hope of
detecting such a process at the Tevatron.

\newpage
{\Large \bf Acknowledgments}
\vspace{1cm}

We would like to thank Alexander Belyaev for contributions at
early stages of this work. We also thank Prof. R. Casalbuoni
for correspondence on the gauge invariance problem.
This work was supported by Conselho Nacional de Desenvolvimento
Cient\'{\i}fico e Tecnol\'ogico (CNPq)
grant {\bf 300538/94-4}, Funda\c{c}\~ao de Amparo \`a
Pesquisa do Estado de S\~ao Paulo (FAPESP)
{\bf 9706112-0}, Programa de Apoio a
N\'ucleos de Excel\^encia (PRONEX) and a C\'atedra Presidencial
1997 (Iv\'an Schmidt)

\newpage
{\Large \bf Figure Captions}
\vspace{2cm}

{\bf Figure 1:} Example of $g-\rho_{T8}$ mixing. There is a
similar diagram with quarks in the initial state.

\vspace{1cm}

{\bf Figure 2:} Diagrams that contribute to technirho decay into
two gluons. Traditionally only a) is taken into account.

\vspace{1cm}

{\bf Figure 3:} Feynman diagrams relevant to the production of a
pair of color-octet technietas.

\vspace{1cm}

{\bf Figure 4:} Expected event distribution for
signal with detector resolution described in the text(solid line),
with detector resolution of
$ \frac{\sigma(E)}{E}=\frac{0.10}{\sqrt{E(GeV)}}$ for photons
and
$  \frac{\sigma(E)}{E}=\frac{0.20}{\sqrt{E(GeV)}}$ for jets
(dotted line)
 and background (dashed line), for $M_{\eta}=150$ GeV and
 $M_{\rho}=410$ GeV. \\

\begin{table}[htbp]
 \caption{Signal total cross section for various technirho and
   technipions masses}
\begin{center}
\begin{tabular}{|c|c|c|c|}
\hline\
$M_{\rho}$ (GeV)&$ M_{\eta}$ (GeV)& $\Gamma_{\rho}$
(GeV)&$\sigma_{\mbox{\tiny{signal}}}$  $(\pb)$ \\
\hline
400&150&77.4&0.0466\\
600&150&250.0&0.0172\\
600&250&71.7&0.0066\\
800&150&407&0.0015\\
800&250&248&0.0124\\
\hline
\end{tabular}
\end{center}
\end{table}

\begin{figure}[htbp]
  \begin{center}
    \begin{picture}(400,100)(0,0)
    \Gluon(100,75)(150,50){4}{5}
    \Gluon(150,50)(100,25){4}{5}
    \Gluon(150,50)(200,50){4}{5}
    \SetWidth{5}
    \Line(200,50)(250,50)
    \SetWidth{0.5}
    \DashLine(250,50)(300,75){4}
    \DashLine(250,50)(300,25){4}
     \Text(125,75)[]{$g$}
     \Text(125,25)[]{$g$}
     \Text(175,60)[]{$g$}
     \Text(225,60)[]{$\rho_T$}
     \Text(320,75)[]{$\eta_T$}
     \Text(320,25)[]{$\eta_T$}
    \end{picture}
    \vskip0.5cm
    \caption{Example of $g-\rho_{T8}$ mixing. There is a similar diagram with
quarks in the initial state.}
    \label{fig:vmd}
  \end{center}
      \vskip0.5cm

  \begin{center}
    \begin{picture}(150,100)(0,0)
     \SetWidth{5}
     \Line(0,50)(50,50)
     \SetWidth{0.5}
     \Gluon(50,50)(100,50){4}{5}
     \Gluon(100,50)(150,100){4}{5}
     \Gluon(100,50)(150,0){-4}{5}
     \Text(75,0)[]{a)}
    \end{picture}
\hspace{1.0cm}
    \begin{picture}(150,100)(0,0)
     \SetWidth{5}
     \Line(0,50)(50,50)
     \Line(50,50)(100,75)
     \Line(50,50)(100,25)
     \SetWidth{0.5}
     \Gluon(100,75)(150,100){4}{5}
     \Gluon(100,25)(150,0){-4}{5}
     \Text(75,0)[]{b)}
    \end{picture}
    \vskip0.5cm
    \caption{Diagrams that contribute to technirho decay into two
     gluons. Traditionally only a) is taken into account.}
    \label{fig:rhodecay}
  \end{center}
      \vskip0.5cm

  \begin{center}
    \begin{picture}(95,79)(0,0)
      \Text(15.0,70.0)[r]{$g$}
      \Gluon(16.0,70.0)(37.0,60.0){3}{3}
      \Text(15.0,50.0)[r]{$g$}
      \Gluon(16.0,50.0)(37.0,60.0){-3}{3}
      \Text(47.0,70.0)[b]{$g$}
      \Gluon(37.0,60.0)(58.0,60.0){3}{2.5}
      \Text(80.0,70.0)[l]{$\eta_T$}
      \DashLine(58.0,60.0)(79.0,70.0){4.0}
      \Text(80.0,50.0)[l]{$\eta_T$}
      \DashLine(58.0,60.0)(79.0,50.0){4.0}
      \Text(47,20)[b] {diagr.1}
    \end{picture}
    \begin{picture}(95,79)(0,0)
      \Text(15.0,70.0)[r]{$g$}
      \Gluon(16.0,70.0)(58.0,70.0){3.0}{5}
      \Text(80.0,70.0)[l]{$\eta_T$}
      \DashLine(58.0,70.0)(79.0,70.0){4.0}
      \Text(57.0,60.0)[r]{$\eta_T$}
      \DashLine(58.0,70.0)(58.0,50.0){4.0}
      \Text(15.0,50.0)[r]{$g$}
      \Gluon(16.0,50.0)(58.0,50.0){-3.0}{5}
      \Text(80.0,50.0)[l]{$\eta_T$}
      \DashLine(58.0,50.0)(79.0,50.0){4.0}
      \Text(47,20)[b] {diagr.2}
    \end{picture}
    \begin{picture}(95,79)(0,0)
      \Gluon(16.0,70.0)(47.5,60){3}{3}
      \Gluon(16.0,50.0)(47.5,60){-3}{3}
      \DashLine(47.5,60)(79.0,70.0){4.0}
      \DashLine(47.5,60)(79.0,50.0){4.0}
      \Text(15.0,70.0)[r]{$g$}
      \Text(15.0,50.0)[r]{$g$}
      \Text(80.0,70.0)[l]{$\eta_T$}
      \Text(80.0,50.0)[l]{$\eta_T$}
      \Text(47,20)[b] {diagr.3}
    \end{picture}
    \begin{picture}(95,79)(0,0)
      \Text(15.0,70.0)[r]{$q$}
      \ArrowLine(16.0,70.0)(37.0,60.0)
      \Text(15.0,50.0)[r]{$\bar{q}$}
      \ArrowLine(37.0,60.0)(16.0,50.0)
      \SetWidth{5}
      \Line(37,60)(58,60)
      \SetWidth{0.5}
      \Text(80.0,70.0)[l]{$\eta_T$}
      \DashLine(58.0,60.0)(79.0,70.0){4.0}
      \Text(80.0,50.0)[l]{$\eta_T$}
      \DashLine(58.0,60.0)(79.0,50.0){4.0}
      \Text(47,20)[b] {diagr.4}
      \Text(47,75)[]{$\rho^0_8$}
    \end{picture}
    \begin{picture}(95,79)(0,0)
      \Text(15.0,70.0)[r]{$q$}
      \ArrowLine(16.0,70.0)(37.0,60.0)
      \Text(15.0,50.0)[r]{$\bar{q}$}
      \ArrowLine(37.0,60.0)(16.0,50.0)
      \Gluon(37,60)(58,60){4}{2}
      \Text(80.0,70.0)[l]{$\eta_T$}
      \DashLine(58.0,60.0)(79.0,70.0){4.0}
      \Text(80.0,50.0)[l]{$\eta_T$}
      \DashLine(58.0,60.0)(79.0,50.0){4.0}
      \Text(47,20)[b] {diagr.5}
      \Text(47,75)[]{$g$}
    \end{picture}
    \vskip0.5cm
    \caption{Feynman diagrams relevant to the production of a pair of
      color-octet technietas.}
    \label{fig:2eta}
  \end{center}
\end{figure}

\newpage
\begin{figure*}[t]
\begin{center}
\epsfig{file=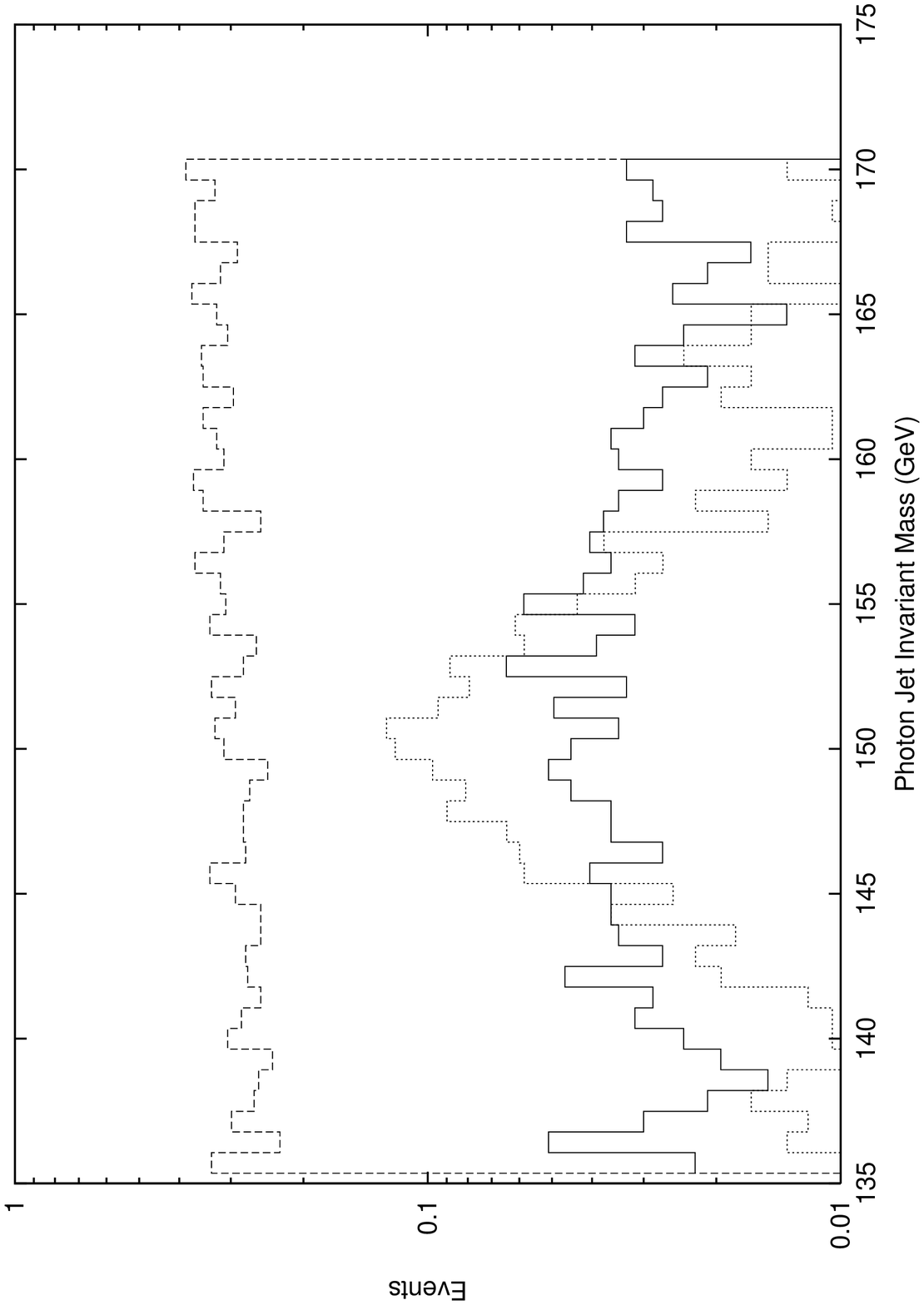,angle=0,width=15cm,height=20cm} \@
\end{center}
\end{figure*}


\newpage
\begin{thebibliography}{99}
\bibitem{Altarelli} For a review, see G. Altarelli, {\tt hep-ph/0011078},
\bibitem{dsb}For a collection of reprints, see
{\sl Dynamical Gauge Symmetry Breaking}, edited by E.~Farhi and
R.~Jackiw
(World Scientific, 1982). More recent developments can be found
in:
K.~Lane, lectures at the LNF Spring School in Nuclear, Subnuclear
 and Astroparticle Physics, Frascati (Rome), Italy, May 15-20,
 2000-{\tt hep-ph/0007304}; 
R.~S.~Chivukula, lectures presented at TASI 2000, Flavor Physics
for the
Millennium, June 4-30, 2000, University of Colorado, Boulder, CO -
{\tt hep-ph/0011264}
R.~S.~Chivukula, R.~Rosenfeld, E.~H.~Simmons and J.~Terning,
in {\sl Electroweak Symmetry Breaking and New Physics at the TeV
Scale},
edited by T.~L.~Barklow, S.~Dawson, H.~E.~Haber and J.~L.~Siegrist
(World Scientific, 1996).

\bibitem{us} A.~Belyaev, R.~Rosenfeld and A.~R.~Zerwekh, \PLB{462}{99}{150}.

\bibitem{EHLQ}E.~Eichten, I.~Hinchliffe, K.~Lane and C.~Quigg,
\RMP{56}{84}{579}.

\bibitem{lr} K.~Lane and  M.~V.~Ramana, \PRD{44}{91}{2678}.


\bibitem{MAD}T.~Stelzer and W.~F.~Long, {\em Comput. Phys. Commun.}
{\bf 81} (1994) 357.

\bibitem{HELAS}H.~Murayama, I.~Watanabe and K.~Hagiwara, KEK report
No.~91-11, unpublished.

\bibitem{CTEQ}H.~L.~Lai et al., CTEQ Collab., \PRD{55}{97}{1280}.





\end{thebibliography}
\end{document}